\documentclass[aps,prd,twocolumn,preprintnumbers,superscriptaddress,nofootinbib,floatfix]{revtex4-1}
\usepackage{graphicx,epstopdf}
\usepackage{amsmath,amssymb,graphicx}
\usepackage[usenames, dvipsnames]{color}
\usepackage{times}
\usepackage{array}
\usepackage{latexsym}
\usepackage{floatrow}
\usepackage{longtable}
\usepackage[utf8, latin9]{inputenc}
\usepackage{tikz,hyperref}
\definecolor{steelblue}{RGB}{25,25,112}
\hypersetup{colorlinks,linkcolor={blue},citecolor={blue},urlcolor={steelblue}}

\definecolor{darkyellow}{rgb}{0.545098,0,0}
\definecolor{teal}{rgb}{0,0.5,0.5}

\newcommand{\hkpc}{h^{-1} ~ {\rm kpc}}
\newcommand{\hMsun}{h^{-1} M_\odot}
\newcommand{\Msun}{M_\odot}
\newcommand{\kmps}{{\rm km \, s^{-1}}}

\newcommand{\MH}{M_{\rm H}}
\newcommand{\rH}{r_{\rm H}}
\newcommand{\vH}{v_{\rm H}}
\newcommand{\nf}{n_{\rm field}}
\newcommand{\sigf}{\sigma_{\rm field}}

\newcommand{\Mvir}{M_{\rm vir}}
\newcommand{\rvir}{r_{\rm vir}}
\newcommand{\Vmax}{V_{\rm max}}
\newcommand{\rmax}{r_{\rm max}}
\newcommand{\rcut}{r_{\rm cut}}
\newcommand{\Mcy}{M_{\rm cy}}

\newcommand{\orcid}[1]{\href{https://orcid.org/#1}{\includegraphics[width=8pt]{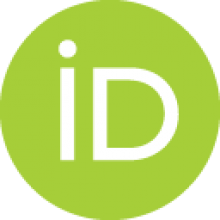}}}

\begin{document}

\title{Searching for Dense Perturbers in Cold Dark Matter Simulations}

\author{Xingyu Zhang\orcid{0009-0009-2791-1684}}
\email{xingyuz@ucr.edu}
\affiliation{Center for Experimental Cosmology and Instrumentation, Department of Physics and Astronomy, University of California, Riverside, California 92521, USA}

\author{Hai-Bo Yu\orcid{0000-0002-8421-8597}}
\email{haiboyu@ucr.edu}
\affiliation{Center for Experimental Cosmology and Instrumentation, Department of Physics and Astronomy, University of California, Riverside, California 92521, USA}

\author{Demao Kong\orcid{0000-0003-1723-8691}}
\email{dkong012@ucr.edu}
\affiliation{Center for Experimental Cosmology and Instrumentation, Department of Physics and Astronomy, University of California, Riverside, California 92521, USA}

\author{Ethan O.~Nadler\orcid{0000-0002-1182-3825}}
\email{enadler@ucsd.edu}
\affiliation{Department of Astronomy \& Astrophysics, University of California, San Diego, California 92093, USA}

\begin{abstract}

Recently, several low-mass, dense perturbers have been inferred from astrophysical observations of gravitational lensing, stellar streams, and satellite galaxies. If interpreted as dark matter subhalos, their inner densities exceed the predictions of the standard cold dark matter (CDM) model. We systematically investigate this tension using a CDM-only zoom-in simulation of a Milky Way analog. We analyze more than $5{,}000$ subhalos with masses down to $10^6\,\Msun$ and reconstruct their inner density profiles using a tidal truncation model. The simulated subhalos are then compared with the observed compact perturbers in terms of their density profiles, enclosed masses, concentrations, and projected cylindrical masses. None of the simulated CDM subhalos is sufficiently dense to reproduce the inferred mass distributions of the observed perturbers. We further examine the redshift evolution of subhalo concentrations and derive a fitting function to characterize this evolution. Our findings motivate further investigations into exceptionally rare formation histories that could produce such dense perturbers, as well as the possibility of physics beyond the standard CDM model.

\end{abstract}
\maketitle

\section{Introduction}
\label{sec-intro}

In the standard cold dark matter (CDM) model, cosmic structure forms hierarchically \citep{Kauffmann:1993umz, Tormen:1996fc, Ghigna:1998vn, Zhao:2002rm, Schaye:2014tpa}. Primordial density fluctuations give rise to the first generation of dark matter halos, after which progressively larger halos assemble through mergers and accretion. Structure formation is approximately self-similar~\cite{Yang:2022oqe}, such that halos over a wide range of masses share a common density profile shape, well described by the Navarro--Frenk--White (NFW) profile \citep{Navarro:1995iw,Navarro:1996gj}. This hierarchical assembly process leads to two key statistical relations for dark matter halos. First, a massive host halo is expected to contain a large population of subhalos, with an approximately scale-free subhalo mass function, $dN/dM \propto M^{-\alpha}$ with $\alpha \sim 2$ \citep{Moore:1999nt, Springel:2008cc}. Second, the halo concentration and mass are tightly correlated \citep{Diemer:2014gba, Prada:2011jf, Sanchez-Conde:2013yxa, Ludlow:2013vxa, Dutton:2014xda, Moline:2016pbm, Ishiyama:2020vao, Moline:2021rza}.

Dark matter halos with shallow potential wells are inefficient at retaining and condensing gas, especially after reionization when the ultraviolet background suppresses gas accretion. The characteristic mass scale below which star formation is strongly suppressed is $M_{\rm crit}(z)\sim 2\times 10^6\text{--}10^9\,\Msun$, with details depending on redshift and the assumptions of galaxy formation models \citep{Bovill:2008fe, Okamoto:2008sn, Sawala:2014hqa, Oshea:2015pro, Sawala:2016tlo, Benitez-Llambay:2020zbo, Pereira-Wilson:2022yjj, Nebrin:2023yzm, Nadler:2025mnz}. Halos whose peak masses remain below the relevant threshold are therefore expected to form few or no stars, making them difficult to detect optically.  Intriguingly, even without optical counterparts, low-mass (sub)halos can be inferred through their gravitational effects, including perturbations in stellar streams induced by close encounters \citep{Yoon:2010iy, Erkal:2016ekt, Banik:2018pjp, Banik:2019smi,Hilmi:2024,Adams:2024zhi,Lu:2025qbp,Lu:2026zni}, distortions in strong-lensing images \citep{Dalal:2001fq, Kochanek:2003zc, Vegetti:2009cz, Vegetti:2012mc, Hezaveh:2016ltk, Inoue:2015lma}, Shapiro delays \citep{Siegel:2007fz, Baghram:2011is, Clark:2015sha, Dror:2019twh}, Doppler effects in pulsar timing \citep{Baghram:2011is, Chakrabarti:2025hfk}, and measurements of the phase-space distribution of Milky Way stars \citep{Feldmann:2013hqa, Buschmann:2017ams, Ravi:2018vqd}.

Recently, several astrophysical observations have provided hints of low-mass dark matter subhalos. Ref.~\cite{Price_Whelan_2018} identified a gap--spur feature in the GD-1 stellar stream using \textit{Gaia} data \citep{Gaia:2016zol,Gaia:2018ydn}. This feature was subsequently interpreted as the result of a close encounter with a $10^6\text{--}10^8\,\Msun$ dark matter subhalo having a characteristic scale radius smaller than $30\,{\rm pc}$ \citep{Bonaca:2018fek}. Another compact substructure of $10^6 \, \Msun$ has been proposed in connection with Fornax~6, a star cluster in the Fornax dwarf spheroidal galaxy. Based on the metallicity and spatial distribution of its stars, Ref.~\cite{Penarrubia:2024vms} suggested that a compact dark matter subhalo may have captured field stars from the Fornax dwarf, producing the stellar overdensity observed as Fornax~6. More recently, gravitational lensing observations have indicated a perturber with a mass of order $10^6\,\Msun$ in the JVAS B1938+666 lens system (B1938+666--$\mathcal{V}$) \citep{Powell:2025rmj}. A follow-up analysis \citep{Vegetti:2026mmx} further suggested that this perturber contains an enclosed mass of $4\times10^5\,\Msun$ within the inner $10\,{\rm pc}$, embedded in a more extended density envelope with a total mass of $2\times10^6\,\Msun$.

Although the inferred density profiles of these perturbers are model dependent, they all appear remarkably dense. If interpreted as CDM subhalos, they are all outliers relative to the concentration--mass relation predicted by CDM cosmological simulations \citep{Moline:2016pbm}. For example, the B1938+666--$\mathcal{V}$ perturber is an extreme $\sim9\sigma$ outlier~\citep{Vegetti:2026mmx,Yu:2025tmp}. Explaining the formation of such compact perturbers therefore poses a potential challenge for the standard CDM model. In contrast, dense subhalos are a generic prediction of self-interacting dark matter (SIDM) models with sufficiently large self-interaction cross sections~\citep{Tulin:2017ara}, in which relatively concentrated halos undergo gravothermal collapse, developing central densities substantially higher than those of their CDM counterparts, e.g.,~\cite{Balberg:2002ue,Essig:2018pzq,Nishikawa:2019lsc,Sameie:2019zfo,Zeng:2021ldo}. Ref.~\citep{Yu:2025tmp} proposed that core-collapsed SIDM subhalos could provide a common origin for these dense perturbers, an intriguing scenario supported by numerical simulations~\citep{Zhang:2024fib, Fischer:2025rky, Schmidt:2026hmn, Zhang:2026gur}.

In this work, we perform a more systematic assessment of whether CDM can explain the three dense perturbers. Previous studies have relied on concentration--mass relations extrapolated into the low-mass regime relevant to these systems, rather than directly comparing the perturbers with simulated subhalos. A further challenge is that the characteristic scales of the inferred perturbers are of order $10\,{\rm pc}$, typically smaller than the spatial resolution of the simulations used to calibrate these relations. 

To overcome these challenges, we directly compare the mass models of the perturbers with CDM subhalos identified in a high-resolution zoom-in simulation of a Milky Way analog~\citep{Nadler:2025mnz}. We analyze more than $5{,}000$ subhalos of the host halo and reconstruct their subsequent density profiles from their infall properties, accounting for tidal truncation using the model in~\citep{Errani:2020wgn}. These reconstructed profiles bypass the resolution limitations, enabling detailed comparisons between the perturbers and simulated CDM subhalos in terms of their inner density profiles, enclosed masses, projected cylindrical mass profiles, and concentrations. We show that none of the simulated CDM subhalos are sufficiently dense to reproduce the adopted mass models of the dense perturbers. We further investigate the redshift evolution of subhalo concentrations and derive a fitting function to characterize this evolution.

The rest of the paper is organized as follows. In Sec.~\ref{sec-perturbers}, we discuss the mass models for the three dense perturbers inferred from observations. In Sec.~\ref{sec-data}, we present the simulation data and our reconstruction method. In Sec.~\ref{sec-lens}, we compare the lensing perturber with simulated subhalos at $z=0.89$. In Sec.~\ref{sec-result}, we compare the GD-1 and Fornax perturbers with simulated subhalos at $z=0$. In Sec.~\ref{sec-redshift}, we examine the redshift dependence of subhalo concentrations. Section~\ref{sec-conclusion} presents our conclusions. In Appendix~\ref{sec:app}, we examine the distribution of subhalo concentrations for a $10^{11}\,\Msun$ host halo, further demonstrating that our conclusions are robust to variations in host halo mass.

\section{Perturbers inferred from astrophysical observations}
\label{sec-perturbers}

In this section, we present mass models for the three dense perturbers. For the lensing perturber, B1938+666--$\mathcal{V}$, the inferred mass distribution is highly compact: under the assumption of a pseudo-Jaffe profile, the perturber has a truncation radius of $149\,{\rm pc}$, while nearly $40\%$ of its total mass is enclosed within the projected radius $R<80\,{\rm pc}$~\cite{Powell:2025rmj}. Ref.~\cite{Vegetti:2026mmx} further explored a wide range of mass models and redshift configurations, finding that the preferred explanation is a perturber residing within the lens galaxy. The favored mass models require an unresolved central component containing $4\times 10^5\,\Msun$ within $10\,{\rm pc}$, surrounded by a more extended mass component. 

\begin{figure}[t]
	\centering
	\includegraphics[scale=0.6]{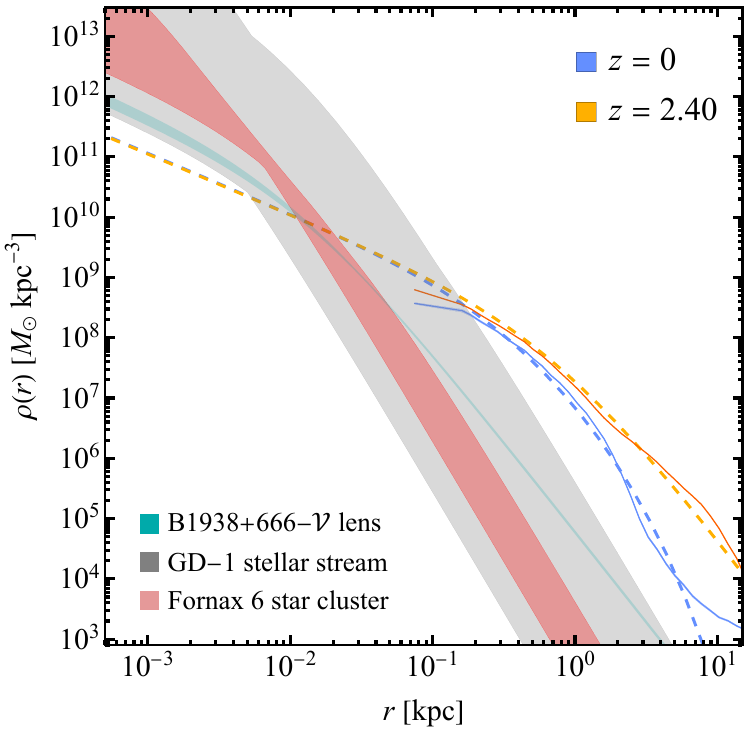}
	\caption{Shaded regions indicate the inferred ranges of density profiles for the B1938+666--$\mathcal{V}$ (cyan), GD-1 (gray), and Fornax (pink) perturbers, as described in Sec.~\ref{sec-perturbers}. Solid curves show simulated density profiles for a representative MW004 subhalo at infall ($z=2.4$, orange) and at $z=0$ (blue), while dashed curves show the corresponding anchored fits described in Sec.~\ref{sec-data}. }
	\label{fig-uplimit}
\end{figure}

For comparison with our CDM simulation, we follow Ref.~\cite{Vegetti:2026mmx} and adopt the NFW mass model for B1938+666--$\mathcal{V}$. 
\begin{equation}
	\rho_{\rm NFW}(r) = \frac{\rho_s}{\left( r/r_s \right)\left( 1 + r/r_s \right)^2},
\end{equation}
where $r_s$ and $\rho_s$ are the scale radius and density, respectively. To determine their values, we adopt $\rmax = 24\pm 4\,{\rm pc}$ for the radius of maximum circular velocity and $M(<\rmax) = (3.6 \pm 0.3)\times10^5 \, \Msun$ for the mass enclosed within $\rmax$. The $1\sigma$ range of the density profile is shown by the cyan band in Fig.~\ref{fig-uplimit}. The NFW model does not include a central point mass and can therefore be compared directly at the density-profile level. Although it is not the best-fitting model identified in~\cite{Vegetti:2026mmx}, it provides a more conservative and less compact representation of the perturber.

For the perturber of the GD-1 stellar stream, Ref.~\cite{Bonaca:2018fek} assumed a Hernquist profile \citep{Hernquist:1990be}
\begin{equation}
	\rho (r)= \frac{M_{\rm H}}{2 \pi r_{\rm H}^3} \, \frac{1}{\left(r/r_{\rm H}\right)\left( 1+ r/r_{\rm H} \right)^3}
	\label{eq:hernquist}
\end{equation}
for the perturber and inferred its viable region in the mass--scale radius space. The constraints are $M_{\rm H} \sim 10^6 \textup{--} 10^8 \, \Msun$ and $\rH \lesssim 30 \, {\rm pc}$ (see Fig.~6 of Ref.~\cite{Bonaca:2018fek}). We use this inferred viable range to reconstruct the corresponding envelope of density profiles, shown by the gray band in Fig.~\ref{fig-uplimit}.

Fornax~6 is a star cluster that resides within the Fornax dwarf satellite galaxy of the Milky Way. It has a total stellar mass of $M_\star = (7.2 \pm 2.2) \times 10^3\,M_\odot$ \citep{DES:2018jtu, DES:2019qxg} and a velocity dispersion of $\sigma_\star = 5.6^{+2.0}_{-1.6}\,\kmps$, implying, under the assumption of dynamical equilibrium, a mass-to-light ratio ranging from $15$ to $258\,M_\odot/L_\odot$ \citep{Pace:2021spe}. The stellar population of Fornax~6 has a metallicity of ${\rm [Fe/H]} = -0.71\pm0.05$ \citep{Pace:2021spe}, similar to that of the metal-rich stellar population of its host galaxy and indicates an age of $\sim2$ Gyr. This contrasts sharply with the five other stellar clusters in Fornax, whose stellar populations are significantly older. Ref.~\cite{Penarrubia:2024vms} proposed that it may comprise stars captured from its host by a compact perturber of the Fornax dwarf galaxy. Under this scenario, we follow the formulation in ~\cite{Penarrubia:2024vms} to compute the density profile of the perturber associated with Fornax~6. We assume that the perturber follows a Hernquist profile as in Eq.~\ref{eq:hernquist} with mass $\MH$ and scale radius $\rH$, and moves at a speed $\vH$ through a background of field stars with number density $\nf$ and velocity dispersion $\sigf$. The number of stars that are temporarily bound to this perturber is
\begin{equation}
	\label{eq-NbF6}
	\begin{aligned}
		N_b(r)	= & \frac{32 \sqrt{\pi}}{9}\left(G \MH \right)^{3 / 2} e^{-\vH^2 /\left(2 \sigf^2\right)} \frac{\nf}{\sigf^3} \\
		& \times \frac{r^2-4 \rH r+8 \rH^{3 / 2}\left( r + \rH \right)^{1 / 2} - 8 \rH^2}{\left(r+\rH\right)^{1 / 2}},
	\end{aligned}
\end{equation}
and the overdensity relative to the field-star background is 
\begin{equation}
	\label{eq-deltabF6}
	\begin{aligned}
		\delta_b(r) & \equiv  \frac{n_b(r)}{\nf}\\
		& =\frac{4}{3 \sqrt{\pi}} \frac{\left(G \MH \right)^{3 / 2}}{\sigf^3} e^{-\vH^2 /\left(2 \sigf^2\right)} \frac{1}{\left(r+\rH\right)^{3 / 2}}.
	\end{aligned}
\end{equation}
By requiring that $\delta_b > 1$ and taking the limit $\rH \rightarrow 0$, we obtain the ``thermal critical radius'':
\begin{equation}
	\label{eq-repF6}
	r_{\epsilon} = \left( \frac{16}{9\pi} \right)^{1/3} e^{-\vH^2/\left( 3\sigf^2 \right)} \frac{G\MH}{\sigf^2},
\end{equation}
while a non-zero $\rH$ gives a radius $r_{\epsilon}^{\rm ext} = r_{\epsilon} - \rH$ within which $\delta_b > 1$. 

The metal-rich subgroup of the Fornax dwarf galaxy contains $N_\star^{\rm MR} = 3 \times 10^7$ stars, with a half-light radius of $r_{1/2} = 530\,{\rm pc}$ and velocity dispersion $\sigf = 10\,\kmps$ \citep{Penarrubia:2024vms}. Assuming a constant number density within $r_{1/2}$, we obtain
\begin{equation}
	\nf = (N_{\star}^{\rm MR}/2)/(4\pi r_{1/2}^3/3) = 2.4\times10^7\,{\rm kpc^{-3}}.
\end{equation}
Substituting these values into Eq.~\ref{eq-NbF6} and assuming that the perturber associated with Fornax~6 moves at the speed $\vH = \sigf$ relative to its host, we obtain the number of stars in the overdense region, $N_b(r_\epsilon^{\rm ext})$, which should be equal to the observed number of stars in Fornax~6. For a specific combination of $(N_b(r_\epsilon^{\rm ext}), \, \MH)$, we can solve for $\rH$ and thus obtain the density profile. For the reconstruction, we adopt the ranges $(5.0\text{--}9.4)\times10^3$ and $(15\text{--}258)\times7.2\times10^3\,\Msun$ for $N_b(r_\epsilon^{\rm ext})$ and $\MH$, respectively. These choices constrain the scale radius to $\rH<15\,{\rm pc}$. The resulting range of allowed density profiles is shown by the pink band in Fig.~\ref{fig-uplimit}. We additionally impose a lower bound of $\rH>1\,{\rm pc}$. Although the perturber could, in principle, be arbitrarily compact, we do not consider such extreme cases here. Our comparison therefore adopts a less compact representation of this perturber.

\section{CDM Subhalos}
\label{sec-data}

We describe the CDM simulation used to search for analogs of the dense perturbers and present our method for reconstructing the density profiles of the simulated subhalos. We analyze a CDM-only cosmological zoom-in simulation of a Milky Way analog (MW004) with virial mass $\Mvir = 7.2 \times 10^{11}\,\hMsun$ and radius $\rvir = 183.9\,\hkpc$ at $z=0$, first presented at this resolution in~\cite{Nadler:2025mnz}, which was drawn from the Milky Way-est suite~\cite{Buch:2024ssx}. The particle mass resolution is $M_{\rm part} = 4.4 \times 10^3\,\hMsun$ and the Plummer-equivalent softening length is $\epsilon = 0.057\,{\rm kpc}$. The corresponding spline length is $\ell = 2.8\epsilon = 0.16\,{\rm kpc}$, which characterizes the smoothing scale of gravitational forces between particles \citep{Springel:2000yr, Springel:2005mi}. The adopted cosmological parameters are $h = 0.7$, $\Omega_{\rm m} = 0.286$, and $\Omega_\Lambda = 0.714$. We select subhalos with masses larger than $10^6\,\hMsun$ ($>227$ particles) at $z=0$, and identify their progenitors at infall using the particle-tracking-based finder \textit{Symfind} \citep{Mansfield:2023prs}. In total, there are $5080$ subhalos in MW004 above this threshold at $z=0$. We also analyze a Large Magellanic Cloud analog simulated at the same resolution in the Concerto suite \cite{Nadler:2025jwh} (LMC104), which is drawn from the Symphony suite \cite{Nadler:2022dvo}. The host halo mass is $6.4\times10^{10}\,\hMsun$ and it contains  $436$ subhalos at $z=0$ after applying the same selection cut. In Appendix~\ref{sec:app}, we will present the distribution of LMC104 subhalo concentrations and demonstrate that our conclusions are robust to variations in host halo mass.

To overcome the resolution limitation and reconstruct the inner density profiles of these simulated CDM subhalos, we apply the tidal truncation model developed in~\cite{Errani:2020wgn}. The evolution of a CDM subhalo in a host potential follows the tidal track in the $\rmax\text{--}\Vmax$ plane, and the density profile during tidal stripping can be fitted by the truncated NFW (tNFW) profile 
\begin{equation}
\label{eq-tNFW}
\rho_{\rm tNFW}(r) = \rho_{\rm NFW}(r) \times \frac{{\rm e}^{-r/\rcut}}{\left( 1+r_s/\rcut \right)^{0.3}},
\end{equation}
where $\rho_{\rm NFW}(r)$ is the NFW density profile of the progenitor halo before tidal stripping. Note that once $\rho_s$ and $r_s$ are specified for a progenitor halo, Eq.~\ref{eq-tNFW} contains only one free parameter, $\rcut$, which characterizes the effect of tidal stripping. Ref.~\cite{Errani:2020wgn} showed that $\rcut$ is, in turn, determined by the degree of tidal mass loss. In this work, however, we do not impose this relation explicitly. Instead, we treat $\rcut$ as a free parameter and determine it by fitting Eq.~\ref{eq-tNFW} to the simulated density profiles. 

We model the density profile of each subhalo at $z=0$ using the following procedure, which we refer to as the {\it anchored} fit. For each subhalo, we first determine $\rmax$ and $\Vmax$ at its infall snapshot from the simulation data and compute the corresponding NFW parameters, $r_s=r_{\rm max}/2.163$ and $\rho_s=V_{\rm max}^2/(0.581\,G\,r_{\rm max}^2)$. We then determine $\rcut$ by fitting Eq.~\ref{eq-tNFW} to the simulated density profile at later snapshots, anchoring $\rho_s$ and $r_s$ to their infall values. In practice, we allow $\rho_s$ and $r_s$ to vary within $0.1$ dex of their values inferred from $\rmax$ and $\Vmax$ to optimize the fit. We have verified that, for the majority of the simulated CDM subhalos, the best-fit values of $\rcut$ are consistent with the relation between the truncation radius and the degree of tidal mass loss derived by Ref.~\cite{Errani:2020wgn}. We note that some subhalos exhibit prominent tidal tails because we do not explicitly remove unbound particles during the analysis. To prevent these tails from biasing the anchored fits, we exclude data points that deviate significantly from the expected power-law or exponential decline of the density profile. Since our analysis focuses on the inner densities of subhalos, we expect the impact of unbound particles on our conclusions to be negligible.

We also apply the same procedure to reconstruct the density profile of each progenitor halo at infall. Although most progenitors in the CDM simulation considered here are well described by an NFW profile, some require non-negligible truncation, indicating that they have already undergone tidal stripping before infall. We therefore adopt Eq.~\ref{eq-tNFW} as a unified description of the density profiles of both progenitor halos and present-day subhalos.

In Fig.~\ref{fig-uplimit}, we show the fit for a representative subhalo from the MW004 simulation. The orange and blue solid curves denote the density profiles measured directly from the simulation at the infall ($z=2.4$) and $z=0$ snapshots, respectively, while the corresponding dashed curves show the reconstructed profiles obtained from the anchored fit. The reconstructed profiles closely match the simulated ones at both snapshots for radii larger than the spline length, $r\gtrsim\ell=0.16\,{\rm kpc}$, which characterizes the resolution limit. At smaller radii, the simulated profiles exhibit an artificial core, particularly at $z=0$, due to numerical effects, whereas the reconstructed profiles extrapolate to the cuspy inner structure assumed by the model. This reconstruction is essential for assessing the CDM interpretation of the inferred dense perturbers. Directly using the simulated density profiles would underestimate the central densities of subhalos, making the comparison with the observations artificially unfavorable. Although the cosmological simulation employed here has high spatial resolution, resolving the inner ${\cal O}(10)\,{\rm pc}$ of $10^6\text{--}10^8\,\Msun$ halos remains challenging. The anchored fitting procedure provides a physically motivated way to overcome this resolution limitation.

\section{The Lensing Perturber}
\label{sec-lens}

\begin{figure*}[t]
    \centering
    \includegraphics[scale=0.45]{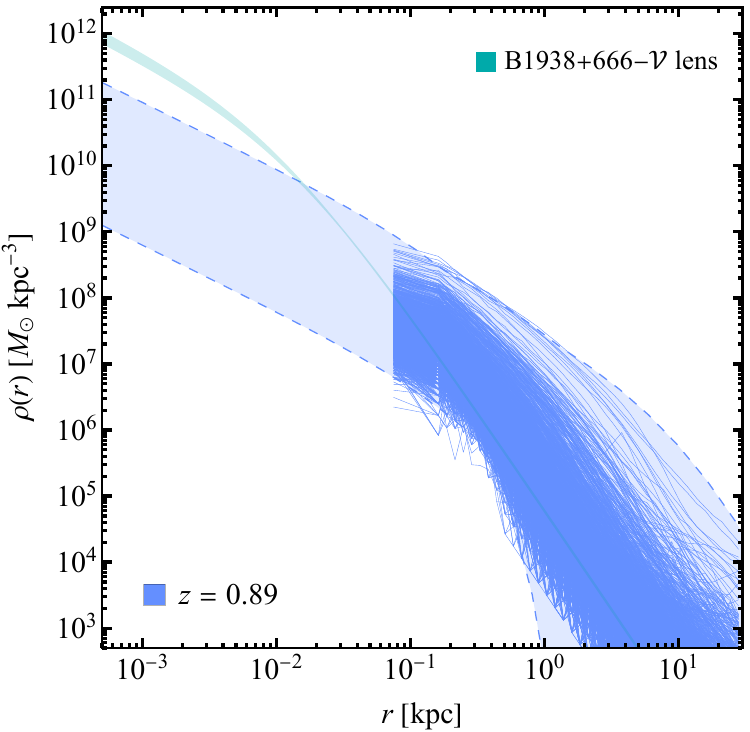}~~~
    \includegraphics[scale=0.45]{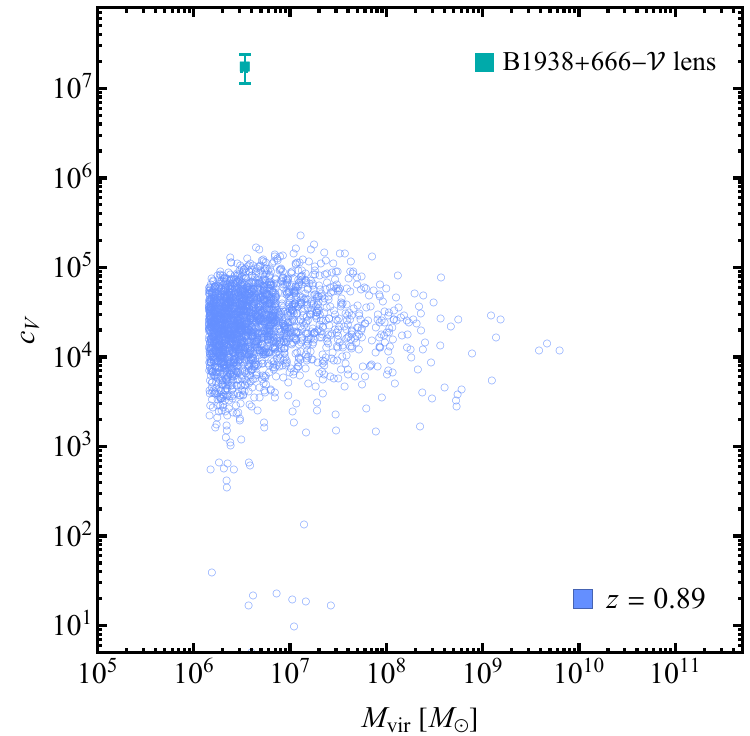}~~~
    \includegraphics[scale=0.45]{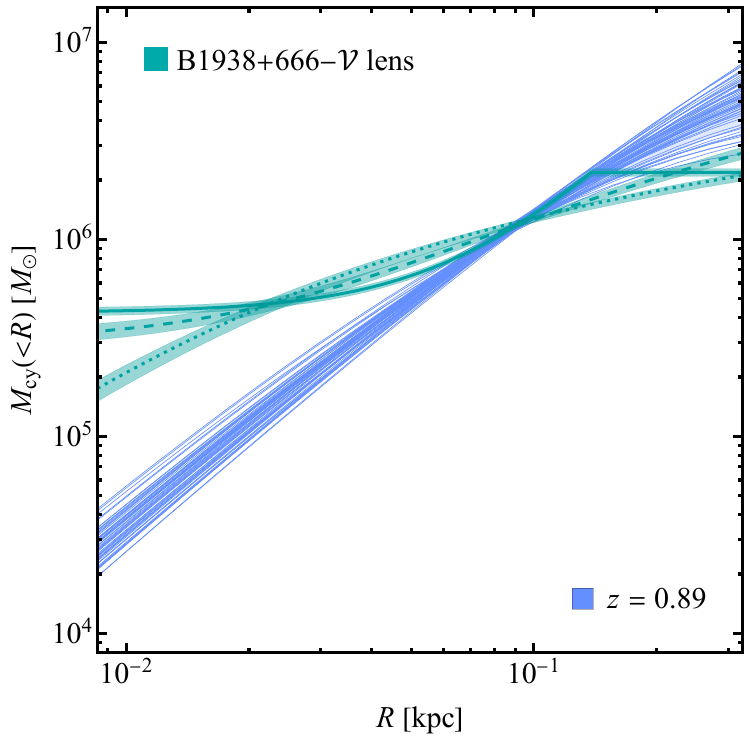}
    \caption{\textbf{Left panel:} Density profiles of MW004 subhalos at $z=0.89$. Solid blue curves show the simulated profiles, while the blue shaded region with dashed boundaries shows the envelope of their anchored fits. The cyan band is the inferred $1\sigma$ range for the NFW model of the B1938+666--$\mathcal{V}$ perturber~\citep{Vegetti:2026mmx}. \textbf{Middle panel:} Concentration $c_V$ versus mass $\Mvir$ for simulated subhalos at $z=0.89$ (blue circles). The cyan square and error bar show the NFW model of the lensing perturber, evaluated at its lens redshift. \textbf{Right panel:} Projected cylindrical mass profiles obtained by integrating the fitted density profiles of individual subhalos along the line of sight. Thin blue curves show subhalos selected to satisfy $\Mcy(<90\,{\rm pc})=(1.1\text{--}1.2)\times10^6\,\Msun$ using these reconstructed profiles. Cyan curves and shaded bands show the inferred profiles and $1\sigma$ uncertainties of the uniform disk plus point mass (UD+PM, solid), NFW plus point mass (NFW+PM, dashed), and NFW (dotted) models~\citep{Vegetti:2026mmx}.}
    \label{fig:lens}
\end{figure*}

The lensing perturber B1938+666--$\mathcal{V}$ is favored to reside in its host lens galaxy at $z=0.881$~\cite{Vegetti:2026mmx}. We therefore compare it with MW004 subhalos at $z=0.89$, the simulation snapshot closest to the lens redshift. We note that the host halo of the lens resides in a different cosmological environment and is more massive than the MW004 host halo. However, these differences do not affect our comparison, as the distribution of subhalo concentrations depends only weakly on host halo mass, see Appendix~\ref{sec:app} and Ref.~\cite{Kong:2025sqx}  for a detailed discussion.

Figure~\ref{fig:lens} (left panel) shows the simulated density profiles as solid blue curves and the envelope of the fitted profiles as a shaded region bounded by dashed curves. The cyan band denotes the density profile reconstructed from the NFW model of the perturber. Even after extrapolating the fitted profiles beyond the simulation resolution limit, the CDM subhalo densities at $r\lesssim10\,{\rm pc}$ remain below the inferred range. As discussed in Sec.~\ref{sec-perturbers}, this NFW model excludes a central point mass and provides a conservative, less compact representation of the perturber. None of the reconstructed profiles in this sample reproduce even this conservative estimate of its inner density.

In Fig.~\ref{fig:lens} (middle panel), we show the concentration parameter~\cite{Diemand:2007qr, Diemand:2008in, Springel:2008cc}
\begin{equation}
\label{eq:cV}
c_V \equiv \frac{\overline{\rho}(\rmax)}{\rho_c(z)} = 2\left(\frac{\Vmax}{H(z)\rmax}\right)^2,
\end{equation}
for the simulated subhalos and the NFW model of the lensing perturber. Here $\overline{\rho}(\rmax)$ is the physical mean density within $\rmax$, $\rho_c(z)$ and $H(z)$ are the critical density and Hubble expansion rate at redshift $z$, respectively. Figure~\ref{fig:lens} (middle panel) shows that the concentration inferred for the NFW perturber model is approximately two orders of magnitude higher than the largest simulated values at the corresponding mass scale.

To directly compare with the gravitational imaging results, in Fig.~\ref{fig:lens} (right panel), we show the projected cylindrical mass profiles
\begin{equation}
\label{eq:Mcy}
\Mcy(<R) \equiv \int_0^R 2\pi R'\,{\rm d}R'\int_{-\infty}^{\infty}{\rm d}z\,\rho \left( \sqrt{z^2+R'^2} \right).
\end{equation}
This quantity is particularly relevant for the lensing comparison because gravitational imaging constrains the projected mass distribution rather than the three-dimensional profiles. The cyan curves represent the uniform disk plus point mass (solid), NFW plus point mass (dashed), and NFW (dotted) models presented in~\cite{Vegetti:2026mmx}, with shaded bands indicating their $1\sigma$ uncertainties. Blue curves represent the profiles for simulated subhalos computed from their anchored fits. To match the observational constraint on the perturber, we include only the subhalos for which $\Mcy(<90 \, {\rm pc})$ is consistent with the observed value of $(1.1\text{--}1.2)\times10^6\,\Msun$. This selection fixes the projected mass normalization at the measured radius, so the remaining comparison probes how rapidly the mass accumulates toward smaller projected radii. We find that, within $R \lesssim 50\,{\rm pc}$, the cylindrical masses of the selected CDM subhalos are significantly lower than those inferred for the lensing perturber. The discrepancy is most pronounced when compared with the two models that include a central point mass.

\begin{figure}[t]
	\centering
	\includegraphics[scale=0.55]{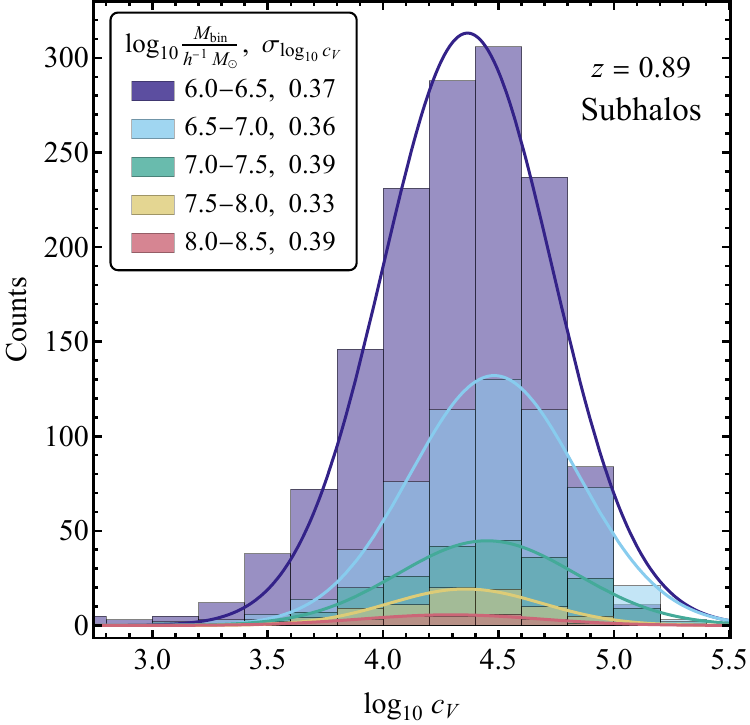}
	\caption{Distributions of $\log_{10}c_V$ for simulated subhalos at $z=0.89$ in the mass bins indicated in the legend. Histograms show the simulation data, and solid curves in matching colors show Gaussian fits. The legend gives the fitted scatter in dex for each bin; masses are expressed in $h^{-1}\Msun$.}
	\label{fig:cv0.89}
\end{figure}

The most intriguing feature of B1938+666--$\mathcal{V}$ is the unresolved ultra-dense inner region within $R=10\,{\rm pc}$, which contains a point mass-like component of $\sim10^5\,\Msun$. The comparison in Fig.~\ref{fig:lens} (right panel) shows that the tension is not simply a deficit in the total projected mass, but rather a deficit in how strongly the mass is concentrated toward the center: even after matching the mass at $R\simeq90\,{\rm pc}$, the CDM subhalos do not place enough mass at smaller radii. A recent study by Ref.~\cite{Zhang:2026gur} showed that such an ultra-dense region can arise naturally in an SIDM scenario through gravothermal collapse, with an effective cross section of $\sigma/m \sim 100\,{\rm cm^2/g}$ in low-mass halos. The central region of a collapsed SIDM halo may even undergo dynamical instability and form a black hole ~\cite{Balberg:2001qg, Feng:2020kxv, Feng:2021rst, Gad-Nasr:2023gvf, Jiang:2025jtr, Feng:2025rzf, Gu:2026zzq, Gu:2026ode}, which accounts for the point-like component of the perturber. An alternative CDM scenario is that the perturber is the remnant of a much more massive progenitor halo of $10^{11}\,\Msun$, whose central density was enhanced by a $10^5\,\Msun$ black hole~\cite{Zhang:2026gur}. A key challenge for this latter scenario is that the progenitor must have lost more than $99.99\%$ of its mass through tidal stripping in the gravitational field of the lens galaxy.

We further quantify the discrepancy of the lensing perturber in the CDM scenario. Figure~\ref{fig:cv0.89} presents the distribution of subhalo concentrations in mass bins spanning $10^6$--$10^{8.5}\,h^{-1}\Msun$. We fit the distribution in each bin with a Gaussian. In the lowest bin, $\Mvir=10^6$--$10^{6.5}\,h^{-1}\Msun$, the median is $\log_{10}c_V=4.37$, with a fitted scatter of $0.37\,{\rm dex}$. Thus, the lensing perturber lies approximately $8\sigma$ above the statistical median. This assessment is based on the NFW model without a central point mass, as shown in Fig.~\ref{fig:lens} (middle panel). The tension is even more severe for the models that include a central point mass, which provide better fits to the lensing imaging data.

\section{The GD-1 and Fornax~6 perturbers}
\label{sec-result} 

\begin{figure*}[t]
	\centering
	\includegraphics[scale=0.45]{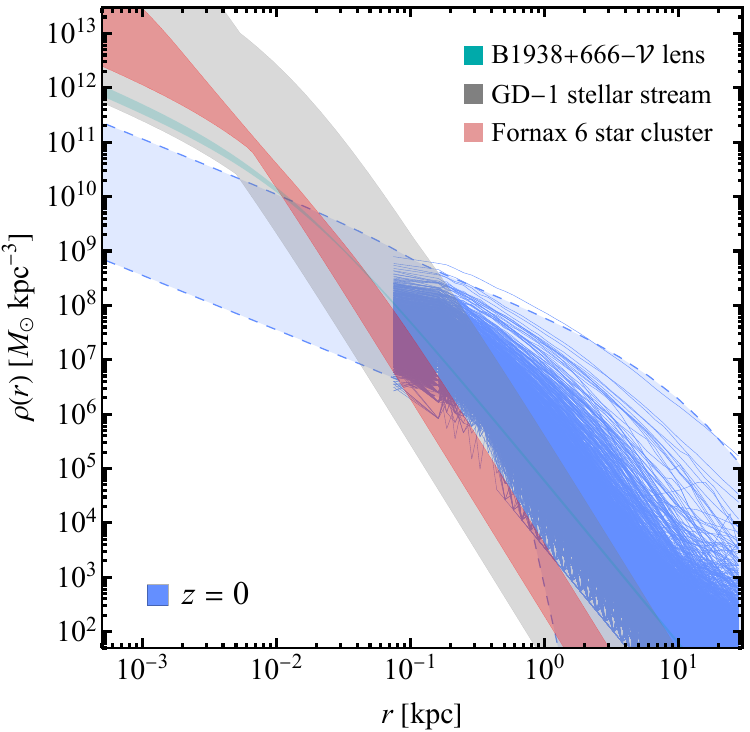}~~~
	\includegraphics[scale=0.45]{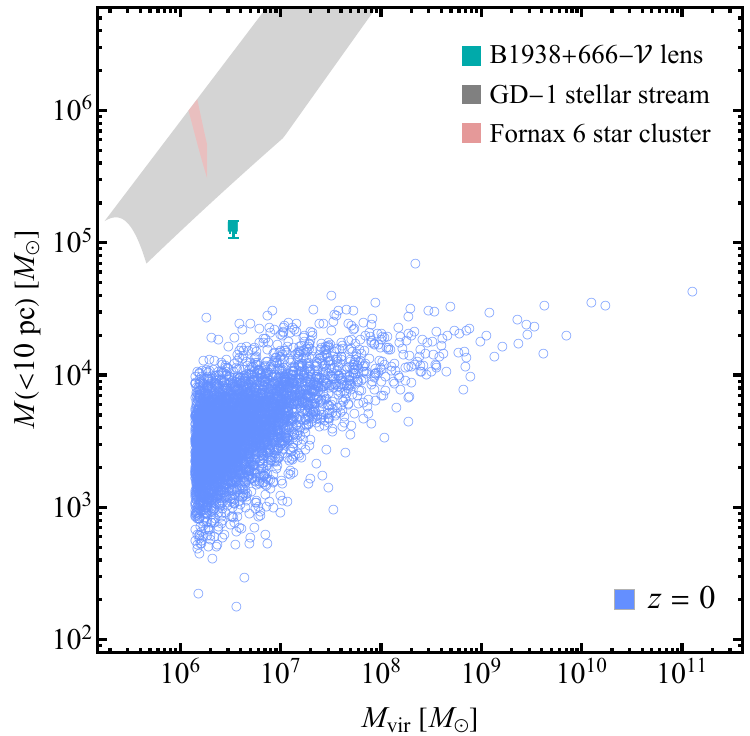}~~~
	\includegraphics[scale=0.45]{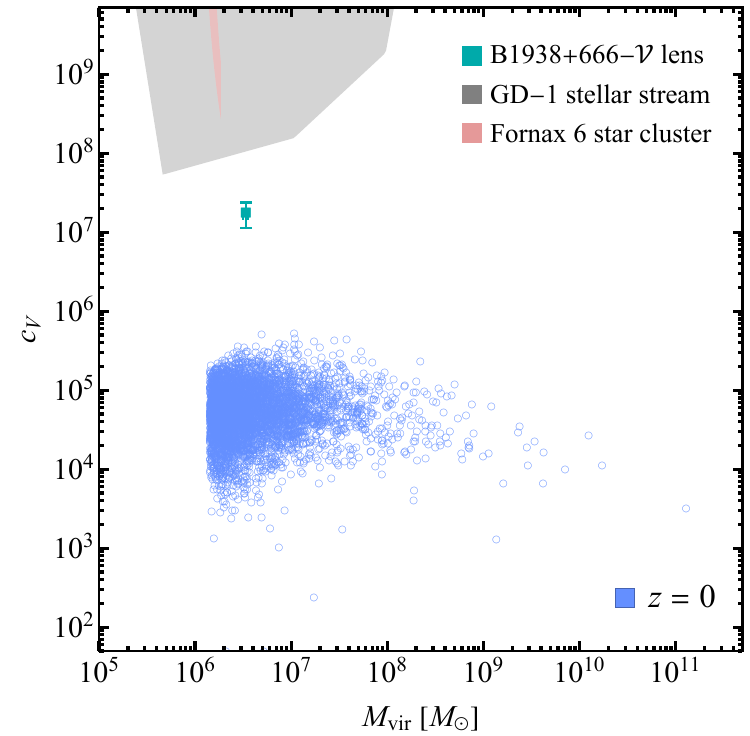}
	\caption{\textbf{Left panel:} Density profiles of MW004 subhalos at $z=0$. Solid blue curves show the simulated profiles, while the blue shaded region with dashed boundaries shows the envelope of their anchored fits. Gray and pink regions show the adopted density-profile ranges for the GD-1 perturber and the perturber associated with Fornax~6, respectively. The cyan band shows the NFW model of the lensing perturber for reference. \textbf{Middle panel:} Enclosed mass within $10\,{\rm pc}$ versus $\Mvir$ for simulated subhalos (blue circles). Gray and pink regions show the corresponding dark matter model ranges for the GD-1 perturber and the perturber proposed for Fornax~6; the cyan square and error bar show the NFW model of the lensing perturber. \textbf{Right panel:} Concentration $c_V$ versus $\Mvir$, with the same color conventions. Note the $c_V$ range for the lensing perturber is evaluated at $z=0.89$ as in Fig.~\ref{fig:lens} (middle panel). }
	\label{fig:GD1F6}
\end{figure*}

Figure~\ref{fig:GD1F6} (left panel) shows simulated density profiles of the CDM subhalos at $z=0$ as solid blue lines and the envelope of the fitted profiles as a shaded region bounded by dashed curves. The favored ranges for the GD-1 and Fornax 6 perturbers reconstructed in Sec.~\ref{sec-perturbers} are shown by shaded regions in gray and pink, respectively. We see that the densities at $r \lesssim 10\,{\rm pc}$ for the simulated CDM subhalos are systematically lower than those of the perturbers. The NFW model of the lensing perturber is included in cyan for reference. 

We also calculate the enclosed mass within a reference radius of $r=10\,{\rm pc}$. This choice is motivated by the inferred compact scales; the GD-1 perturber has a scale radius satisfying the condition $\rH <30 \, {\rm pc}$; the Fornax 6 scenario implies $r_{\rm H}<15\,{\rm pc}$. Thus, $r=10\,{\rm pc}$ probes the central region relevant to the compactness of these systems. Figure~\ref{fig:GD1F6} (middle panel) shows the enclosed mass within $10\,{\rm pc}$ versus the virial mass. The gray and pink regions denote the inferred ranges for the GD-1 and Fornax~6 perturbers, respectively. We see that none of the subhalos are sufficiently dense to match the perturbers. For $M_{\rm vir}\sim10^6\textup{--}10^7\,{M_\odot}$, the $M(<10\,{\rm pc})$ values of the subhalos are at least a factor of $5$ below the viable range of the perturbers.

\begin{figure}[t]
	\centering
	\includegraphics[scale=0.55]{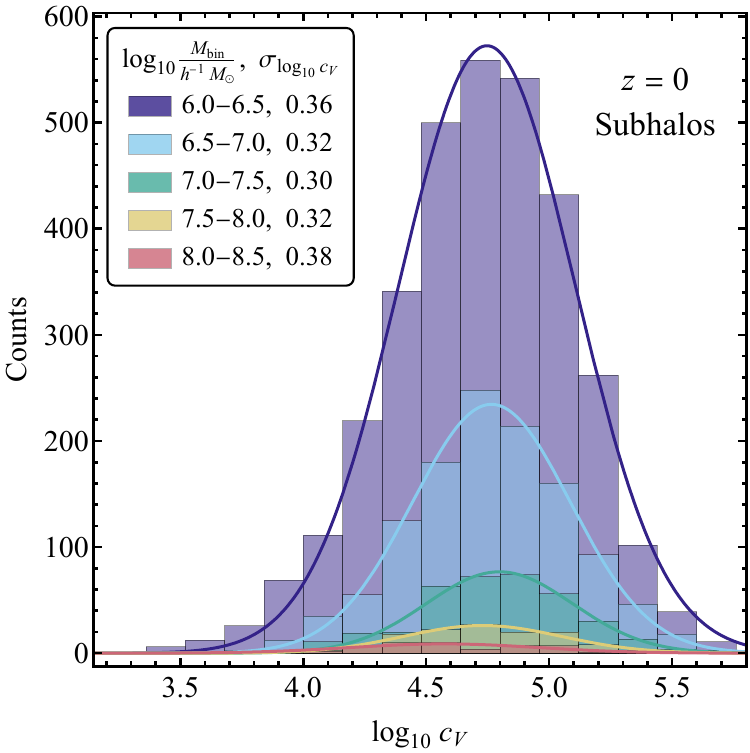}
	\caption{Distributions of $\log_{10}c_V$ for simulated subhalos at $z=0$ in the mass bins indicated in the legend. Histograms show the simulation data, and solid curves in matching colors show Gaussian fits. The legend gives the bin mass and fitted scatter in dex for each bin.}
	\label{fig:cv}
\end{figure}

Figure~\ref{fig:GD1F6} (right panel) shows the concentration--mass distribution at $z=0$, together with the model ranges for the GD-1 and Fornax~6 perturbers. For reference, we also include the NFW model of the lensing perturber evaluated at $z=0.89$. These perturbers have concentrations well above those of the simulated CDM subhalos. Figure~\ref{fig:cv} shows the distributions of $\log_{10}c_V$ for simulated subhalos at $z=0$ and Gaussian fits for individual mass bins. In the lowest bin, $\Mvir=10^6$--$10^{6.5}\,h^{-1}\Msun$, the median is $4.75$ and the fitted scatter is $0.36\,{\rm dex}$. Near this mass range, the lower edge of the GD-1 region is approximately $c_V\sim8\times10^7$, or $\log_{10}c_V\sim7.9$, corresponding to a $9\sigma$ outlier. The Fornax~6 perturber model has even higher concentration, corresponding to a $11\sigma$ outlier.

\section{Redshift dependence of \texorpdfstring{$c_V$}{cV}}
\label{sec-redshift}

\begin{figure*}[t]
	\centering
	\includegraphics[scale=0.45]{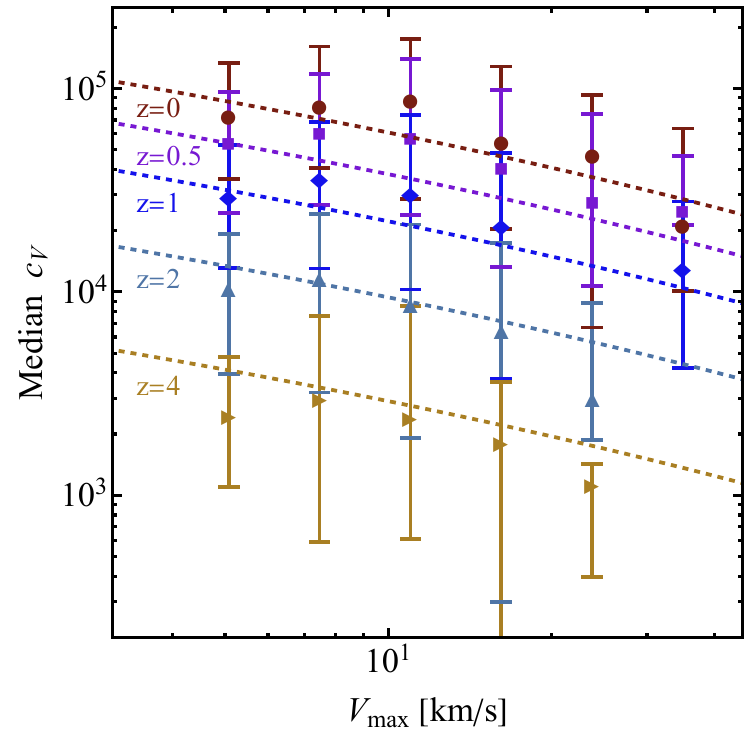}~~~
	\includegraphics[scale=0.45]{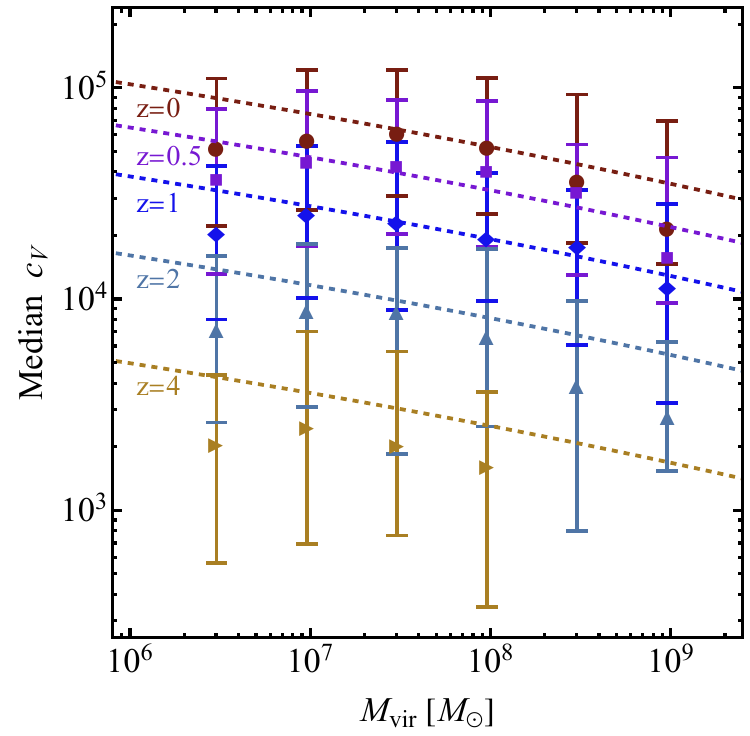}~~~
	\includegraphics[scale=0.45]{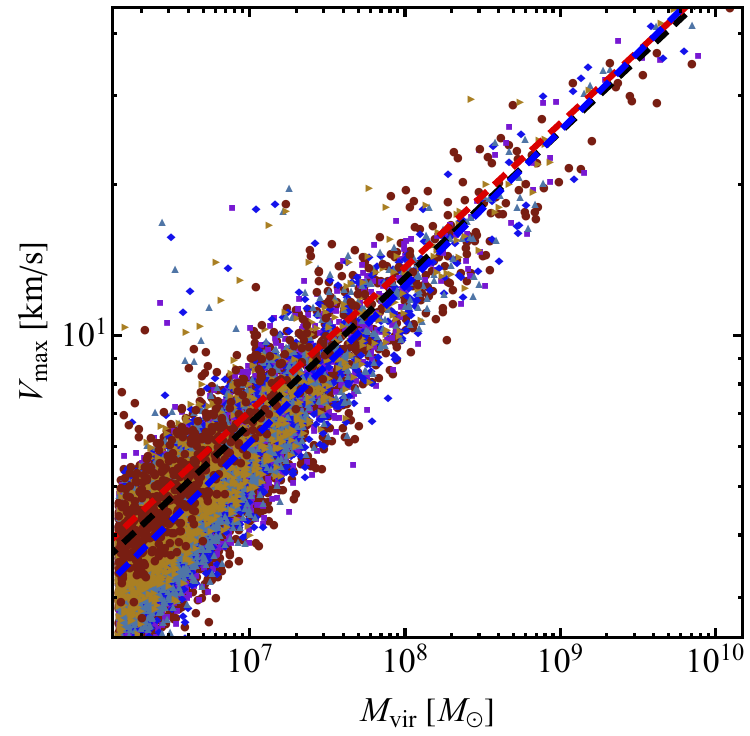}
	\caption{\textbf{Left panel:} Median subhalo concentration $c_V$ as a function of maximum circular velocity $\Vmax$ at $z=0$ (circles), $0.5$ (squares), $1$ (diamonds), $2$ (up-triangles), and $4$ (right-triangles). The scatter spans the 15.9th--84.1st percentiles in each bin. Dashed curves are based on Eq.~\ref{eq:Mo23}. \textbf{Middle panel:} Median concentration as a function of $\Mvir$, with the same colors, symbols, and percentile ranges. Dashed curves are from Eq.~\ref{eq:Mo23}, after using the $V_{\rm max}$--$M_{\rm vir}$ relation in Eq.~\ref{eq:MvirVmax}. \textbf{Right panel:} $\Vmax$ versus $\Mvir$ for individual subhalos at the same redshifts. Dashed curves show the corresponding relations from Eq.~\ref{eq:MvirVmax} (black), Aquarius (red) \cite{Springel:2008cc} and Millennium-II (blue) \cite{Boylan-Kolchin:2009ztl} simulations.
    }
	\label{fig:cVz}
\end{figure*}

In this section, we examine the redshift dependence of subhalo concentrations and compare our results with the parametrization of Ref.~\cite{Moline:2021rza}, a standard reference in the literature. Figure~\ref{fig:cVz} (left panel) shows the median concentration $c_V$ as a function of $\Vmax$ at $z=0$, $0.5$, $1$, $2$, and $4$. The scatter spans the 15.9th--84.1st percentiles of each concentration distribution. At fixed $\Vmax$, the median $c_V$ decreases with increasing redshift by more than an order of magnitude from $z=0$ to $z=4$, while the dependence on $\Vmax$ is comparatively weak, with a general decline toward larger $\Vmax$. To describe these trends, we adopt the functional form proposed in~\cite{Moline:2021rza},
\begin{equation}
	\label{eq:Mo23}
	\begin{aligned}
		c_V(\Vmax,z) = & c_0 \left\{ 1+\sum_{i=1}^{3} a_i \left[ \log_{10}\left(\frac{\Vmax}{\rm km\,s^{-1}}\right)	\right]^i \right\}\\
		&\times (1+z)^b(1+dz).
	\end{aligned}
\end{equation}
We retain the normalization $c_0=1.75\times10^5$, the coefficients $a_1=-0.90368$, $a_2=0.2749$, and $a_3=-0.028$, and $d=3.2$ as in~\cite{Moline:2021rza}, but set $b=-3.52$. The resulting functional fits are shown by dashed curves in Fig.~\ref{fig:cVz} (left panel). Note that Ref.~\cite{Moline:2021rza} set $b=-5.52$, which implies a substantially faster decline in $c_V$ with redshift than we find in our simulation. This discrepancy is consistent with a missing scale factor in the conversion of $\rmax$ from comoving to physical units in~\cite{Moline:2021rza}.

Figure~\ref{fig:cVz} (middle panel) presents the median concentration as a function of $\Mvir$. To express the parametrization in terms of mass, we digitize the correspondence between the upper mass axis and the lower $\Vmax$ axis in Fig.~14 of Ref.~\cite{Moline:2021rza}. This yields the relation
\begin{equation}
  \label{eq:MvirVmax}
  \frac{\Vmax}{\rm km\,s^{-1}} = 13.1\left(\frac{\Mvir}{10^8\,\Msun}\right)^{0.29}.
\end{equation}
Substituting it into Eq.~\ref{eq:Mo23} gives the dashed curves in Fig.~\ref{fig:cVz} (middle panel). We see that the converted $c_V$--$\Mvir$ functional relation also reproduce the overall trends of our simulated concentrations.

Figure~\ref{fig:cVz} (right panel) shows the $\Vmax$--$\Mvir$ distribution of individual subhalos at the same redshifts. The black dashed line shows Eq.~\ref{eq:MvirVmax}, which provides a good fit for the simulation data. It also agrees with the fits from Aquarius~\cite{Springel:2008cc}, $\Vmax = 13.7 (M_{\rm sub} / 10^8 \Msun)^{0.29} \, \kmps$, and Millennium-II~\cite{Boylan-Kolchin:2009ztl}, $\Vmax = 12.5 (M_{\rm sub} / 10^8 \Msun)^{0.31} \, \kmps$, shown by dashed red and blue lines, respectively. We see that our simulated subhalos at different redshifts occupy a broadly similar locus in the $\Vmax$--$\Mvir$ plane. This supports using Eq.~\ref{eq:MvirVmax} as an approximate conversion for the two quantities.

\section{Discussion and Conclusions}
\label{sec-conclusion}

We have investigated whether the standard CDM framework can accommodate the dense perturbers with masses of $\sim 10^6\text{--}10^7\, \Msun$ inferred form observations of the GD-1 stellar stream, the JVAS B1938+666 lens system, and the Fornax~6 stellar cluster. We analyzed more than $5{,}000$ simulated subhalos of a Milky Way analog and reconstructed their density profiles using a tidal truncation model, which provides a physically-motivated extrapolation to the ${\cal O}(10 \,{\rm pc})$ scales relevant to these observations. None of the simulated subhalos are sufficiently dense to reproduce the adopted mass models of the perturbers, making them extreme outliers in concentration within the standard CDM framework.

On the other hand, although the three perturbers reside in different galactic environments and span different cosmic epochs, they exhibit similar masses and concentrations, suggesting that their high densities may arise from a common mechanism, such as SIDM gravothermal collapse~\cite{Yu:2025tmp}. This mechanism has also been proposed to explain other lensing observations, including the high densities of more massive lensing perturbers~\cite{Nadler:2023nrd,Tajalli:2025qjx,Li:2025kpb,Kong:2025sqx,Li:2026duw}, flux-ratio anomalies~\cite{Gilman:2022ida,Gilman:2026wul}, and the excess of small-scale gravitational lenses in galaxy clusters~\cite{Yang:2021kdf,Dutra:2024qac,Ragagnin:2024deh,Yang:2025xsp}. Importantly, SIDM gravothermal collapse can remain robust even in the presence of strong baryonic feedback~\cite{Kong:2026piq,Feng:2025rzf}.

Further work is nevertheless needed to establish the interpretation of these observational ``anomalies'' as dark matter perturbers. For example, a time-dependent Milky Way potential may also explain the gap--spur feature in the GD-1 stream~\cite{Arora_2026}. The lensing perturber could alternatively be a composite system consisting of a central black hole embedded in a massive progenitor halo that has undergone an unusual tidal evolution~\cite{Zhang:2026gur}. Future observations and modeling will be essential to determine whether these observations represent rare astrophysical outliers or genuine evidence for physics beyond the standard CDM framework.

\appendix

\section{LMC104 simulation}
\label{sec:app}

\begin{figure}
	\centering
	\includegraphics[scale=0.5]{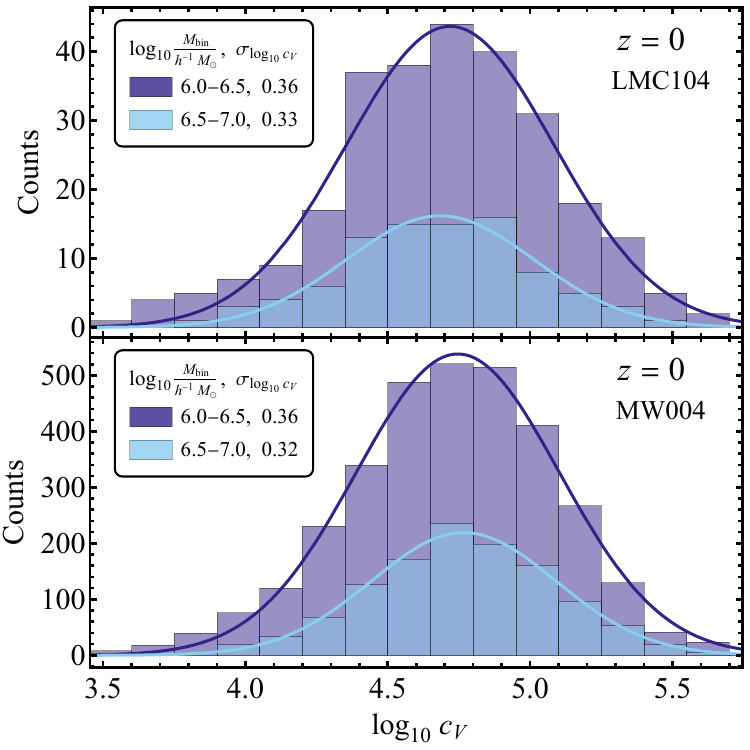}
	\caption{Comparison of $\log_{10}c_V$ distribution between LMC104 (top panel) and MW004 (bottom panel) subhalos at $z=0$. Histograms show the simulation data, and solid curves in matching colors show Gaussian fits. The legend gives the bin mass and fitted scatter in dex for each bin.}
	\label{fig:cv-LMC}
\end{figure}

In this appendix, we present results from the LMC104 simulation of a Large Magellanic Cloud analog host halo with $\Mvir = 6.4 \times 10^{10}\,\hMsun$ and $\rvir = 82.1\,\hkpc$ at $z=0$. The data was first presented in~\cite{Nadler:2022dvo} and later studied as part of the Concerto suite in~\cite{Nadler:2025jwh}. The resolution is the same as that of MW004. At $z=0$, LMC104 contains a total of $436$ subhalos with masses above $10^6\,h^{-1}\Msun$. 

Figure~\ref{fig:cv-LMC} compares the distributions of $\log_{10} c_V$ for LMC104  (top panel) and MW004 (bottom panel) subhalos. We restrict the comparison to the two lowest mass bins to ensure sufficient sample sizes for LMC104. In order of increasing mass, the median values are $4.72$ and $4.68$ for LMC104, compared with $4.75$ and $4.77$ for MW004. The two hosts exhibit similar median values and scatter, suggesting that subhalo concentrations depend only weakly on host halo mass over the range considered.

\section*{Acknowledgments}
This research was supported by the John Templeton Foundation under Grant ID \# 63599 and the U.S. Department of Energy under Grant No. DE-SC0008541. The opinions expressed in this publication are those of the authors and do not necessarily reflect the views of the funding agencies.

\section*{DATA AVAILABILITY} 
The reconstructed density-profile data are publicly available at \cite{data:2026}.

\bibliography{refbib-fitting-new}

\clearpage

\end{document}